\documentclass[letterpaper, 10 pt, conference]{ieeeconf}  % Comment this line out if you need a4paper

\IEEEoverridecommandlockouts                              % This command is only needed if 
\usepackage{graphics} % for pdf, bitmapped graphics files
\usepackage{epsfig} % for postscript graphics files
\usepackage{amsmath} % assumes amsmath package installed
\usepackage{graphicx}
\usepackage{tikz}
\usetikzlibrary{patterns,shapes,arrows,fit,positioning,shapes.geometric,shapes.symbols,shapes.misc,decorations,calc,backgrounds}
\usepackage{pgfplots}
\pgfplotsset{compat=newest}
\pgfplotsset{plot coordinates/math parser=false}
\usepackage{subcaption}
\usepackage{stfloats}
\usepackage{float}
\usepackage{caption}
\usepackage{tikz}
\usepackage{textcomp}
\usepackage{hyperref}
\usepackage{lipsum}

\newlength\figurewidth
\newlength\figureheight

\title{\LARGE \bf
Exploring EEG Responses during Observation of Actions Performed by Human Actor and Humanoid Robot
}

\author{Anh T. Nguyen, \textit{Member, IEEE}$^{1}$, Ajay Anand, \textit{Member, IEEE}$^{2}$ and Michelle J. Johnson, \textit{Member, IEEE}$^{3}$% <-this % stops a space
\thanks{This work was supported by the Department of Physical Medicine and Rehabilitation at the University of Pennsylvania and by the Eunice Kennedy Shriver National Institute of Child Health \& Human Development of the National Institutes of Health (NIH) under Award Number F31HD102165 and the Vingroup Scholarship Program for Master’s and Ph.D. Degrees Overseas Study. The content does not necessarily represent the views of the NIH.}% <-this % stops a space
\thanks{$^{1}$Anh T. Nguyen is with the School of Science and Applied Science, Department of Bioengineering, University of Pennsylvania, Philadelphia, PA, USA
        {\tt\small tuna28ng@seas.upenn.edu}}%
\thanks{$^{2}$ Ajay Anand is with the Rehab Robotics Lab under the General Robotics, Automation, Sensing, \& Perception Lab (GRASP Lab), University of Pennsylvania, Philadelphia, PA, USA
        {\tt\small ajay.anand@pennmedicine.upenn.edu}}%
\thanks{$^{3}$Dr.\ Michelle J. Johnson is an Associate Professor with the Department of Physical Medicine and Rehabilitation and BioEngineering. She directs the Rehab Robotics Lab (A GRASP Lab), University of Pennsylvania, Philadelphia, PA, USA 
        {\tt\small johnmic@pennmedicine.upenn.edu}}%
}

\begin{document}

\maketitle
\thispagestyle{empty}
\pagestyle{empty}

%%%%%%%%%%%%%%%%%%%%%%%%%%%%%%%%%%%%%%%%%%%%%%%%%%%%%%%%%%%%%%%%%%%%%%%%%%%%%%%%
\begin{abstract} % no more than 200 words?

Action observation (AO) therapy is a promising rehabilitative treatment for motor and language function in individuals recovering from neurological conditions, such as stroke. This pilot study aimed to investigate the potential of humanoid robots to support AO therapy in rehabilitation settings. The brain activity of three healthy right-handed participants was monitored with electroencephalography (EEG) while they observed eight different actions performed by two agents, a human actor and a robot, using their left and right arms. Their event-related spectral perturbations (ERSPs, changes in the spectral power of neural oscillations in response to an event or stimulus, compared to baseline) in sensorimotor regions were analyzed. 
The single-subject analysis showed variability in ERSP patterns among all participants, including power suppression in sensorimotor mu and beta rhythms. One participant showed stronger responses to "robot" AO conditions than to "human" conditions. Strong and positive correlations in ERSP across all conditions were observed for almost all participants and channels, implying common cognitive processes or neural networks at play in the mirror neuron system during AO. The results support the feasibility of using EEG to explore differences in neural responses to observation of robot- and human-induced actions.

% Resutls and their significance
% Using humanoid robots can also alleviate some of the burden on the caregivers.
\end{abstract}

%%%%%%%%%%%%%%%%%%%%%%%%%%%%%%%%%%%%%%%%%%%%%%%%%%%%%%%%%%%%%%%%%%%%%%%%%%%%%%%%
\section{INTRODUCTION}

Action observation (AO) therapy is a promising rehabilitative approach rooted in the principles of action perception and execution, designed to harness the inherent neural mechanisms involved in motor learning and recovery.
AO is effective in improving motor functions in the upper extremity and the gait of individuals recovering from neurological conditions, such as stroke, traumatic brain injury, and Parkinson's disease \cite{borges2022action, caligiore2019action}. 
% AO therapy in conjunction with speech language therapy promotes language improvement from post-stroke aphasia and apraxia of speech \cite{murteira2020can, you2019effectiveness}. 
AO therapy is based on understanding the mirror neuron system (MNS), a neural network originally identified in primates that includes the ventral pre-motor area F5 and the inferior parietal area PFG \cite{di1992understanding, gallese1996action, rizzolatti1996premotor, rozzi2008functional, fogassi2005parietal}.
Mirror neurons in visuo-motor areas activate during the observation and execution of a motor act, thus observing actions performed by others produces a motor activation similar to those produced when one is involved in executing those actions. The underlying mechanism is that the observed action is mapped onto the observer's own motor representation, facilitating a comprehension of the action goal \cite{rizzolatti2001neurophysiological}.
Studies on observations of goal-oriented actions in humans showed increased cortical responses in visual areas, as well as in the ventral premotor cortex (PMv) and in several other sensorimotor areas \cite{caspers2010ale, molenberghs2012brain, molenberghs2012brain, filimon2007human, gazzola2009observation}.

% However, the literature also presents the need for further investigations to provide more insights into the ways these factors modulate the action observation system and why AO is effective in rehabilitation. 

A recent review identified multiple factors that influence the patterns and strength of neural responses in the MNS during AO \cite{kemmerer2021modulates}. 
Factors that involve action stimuli such as transitivity and degree of realism and factors involving the actor such as similarity to humans, were identified as having neuromodulation effects on the MNS \cite{kemmerer2021modulates}.
Since the discovery of the Mirror Neuron System (MNS), there has been one question of interest regarding whether artificial agents like robots can activate the MNS and how the responses of the MNS to robotic agents compare to those elicited by human agents.
Addressing this question is crucial for advancing the integration of robots into rehabilitation practices, particularly in AO therapy.
EEG have been used to examine the neural response to robot actions vs. human actions in brain regions involved in AO \cite{oberman2007eeg, urgen2013eeg, gazzola2007anthropomorphic}.
In particular, a well-established EEG pattern referred to as power suppression of mu ($\mu$) band (i.e., a sensorimotor rhythm that occurs in the frequency range of 8 to 13 Hz) is found to be elicited during AO, indicating activation of the mirror neurons \cite{ hobson2017interpretation}. Analyzing mu rhythm suppression, also known as event-related desynchronization (ERD), allows the evaluation of the functional integrity of the MNS. 
An EEG study by Oberman et al. found equivalent mu rhythm (8 13Hz) suppression when a human arm or a robotic arm performed grasping actions \cite{oberman2007eeg}. This result is also consistent with the findings of an fMRI study by Gazzola et al. \cite{gazzola2007anthropomorphic} which found no significant differences in responses to the two types of agents.
The findings of an EEG study by Urgen et al., which examined mu suppression during the observation of humanoid robots with a more human-like appearance, also suggest that human MNS is unlikely to be selective only for the human actor \cite{urgen2013eeg}. 
Cross et al. found that the action observation network can be strongly activated by actions that are unfamiliar to individuals, such as robotic movements, and tends to have a more robust response to robotic movements than to human-like movements \cite{cross2012robotic}.
These results suggest that observing robots can potentially modulate neural activity in regions associated with movement; however, more studies investigating neural responses during observations of actions performed by a humanoid robot compared to a human agent are needed to conclude the efficacy of robot-assisted AO therapy.
To contribute to the existing body of knowledge and lay a foundation for long-term future research exploring the use of social robots in neurorehabilitation, we conducted a case study with three healthy participants undergoing AO tasks with different conditions, using videos featuring a humanoid robot and a human actor as two types of actors, and actions performed by left and right hands.
In contrast to most prior studies, we employed videos of a humanoid robot to match the conventional AO therapy setup, which generally features videos of a human actor.
We collected EEG data during AO and analyzed the event-related spectral perturbation (ERSP) between different AO conditions to explore influence of humanoid robot in AO and verify the feasibility of the EEG-based approach to identify sensorimotor activity.

\section{METHOD}
% \subsection{Participants}
We collected data from three healthy participants, all right-handed as assessed by the Edinburgh Handedness Inventory \cite{oldfield1971assessment}. The participants were between the ages of 21 and 28 years (1 Asian male, 1 black male, and 1 white female) and reported no history of neurological or psychiatric disorders. %(Table \ref{demographics}). 

% \begin{table}[h]
% \caption{Participant demographics} \label{demographics}
% \label{table_example}
% \begin{center}
% \begin{tabular}{|c|c|c|c|c|}
% \hline
% Study ID & Gender & Age & Ethnicity & Dominant Hand \\
% \hline
% Subject 1 & Female & 28 & White & Right \\
% \hline
% Subject 2 & Male & 24 & Black & Right \\
% \hline
% Subject 3 & Male & 21 & Asian & Right \\
% \hline
% \end{tabular}
% \end{center}
% \end{table}

\subsection{Experimental Design}
\begin{figure}[h]
% \centering
\vspace{5pt}
\valign{%
  #\cr
  \hbox{\subcaptionbox* {(A)}[.6\linewidth]{%
    \includegraphics[width=\linewidth]{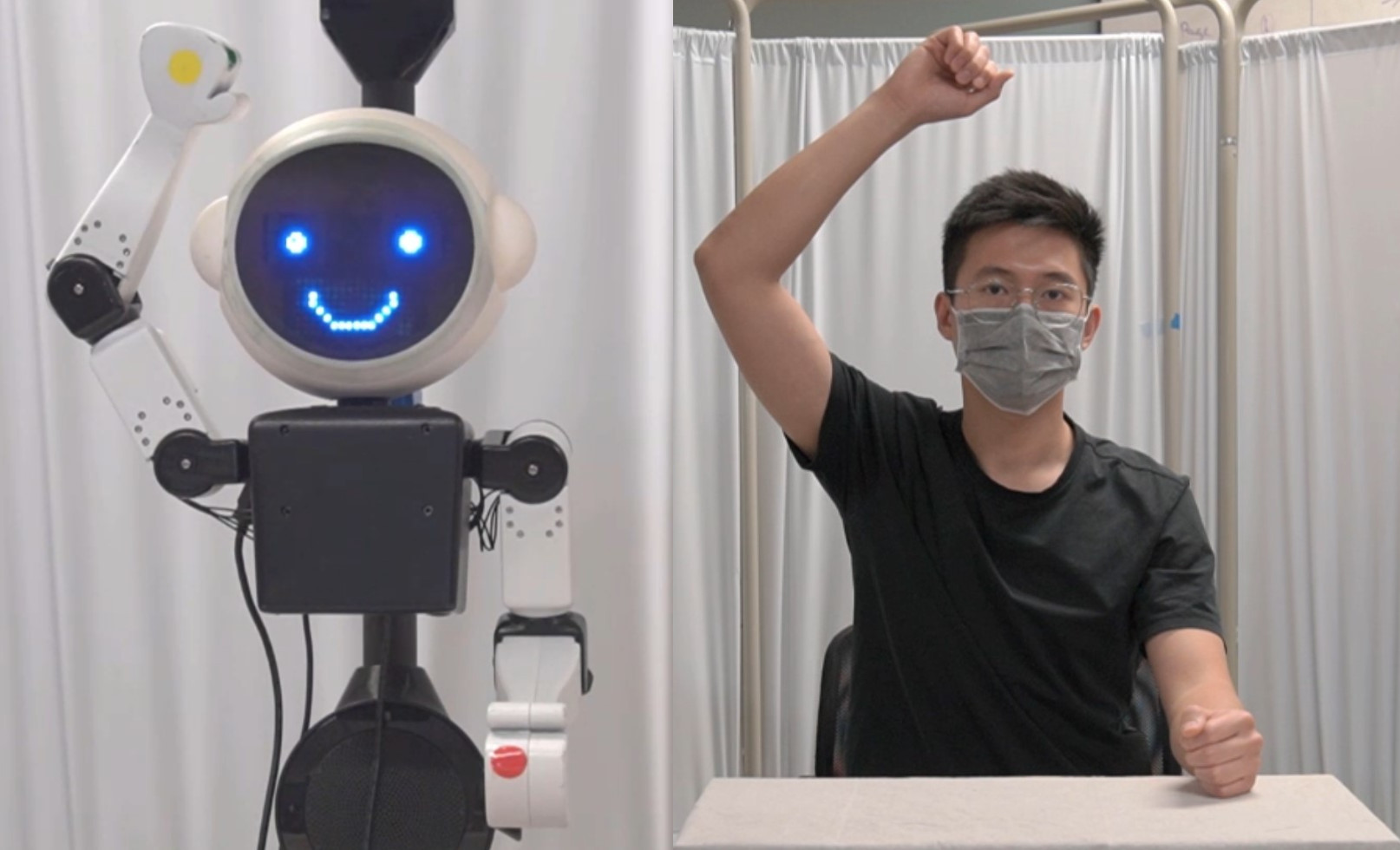}%
  }}
  \vfill
  \hbox{\subcaptionbox*{(B)}[.6\linewidth]{%
    \includegraphics[width=\linewidth]{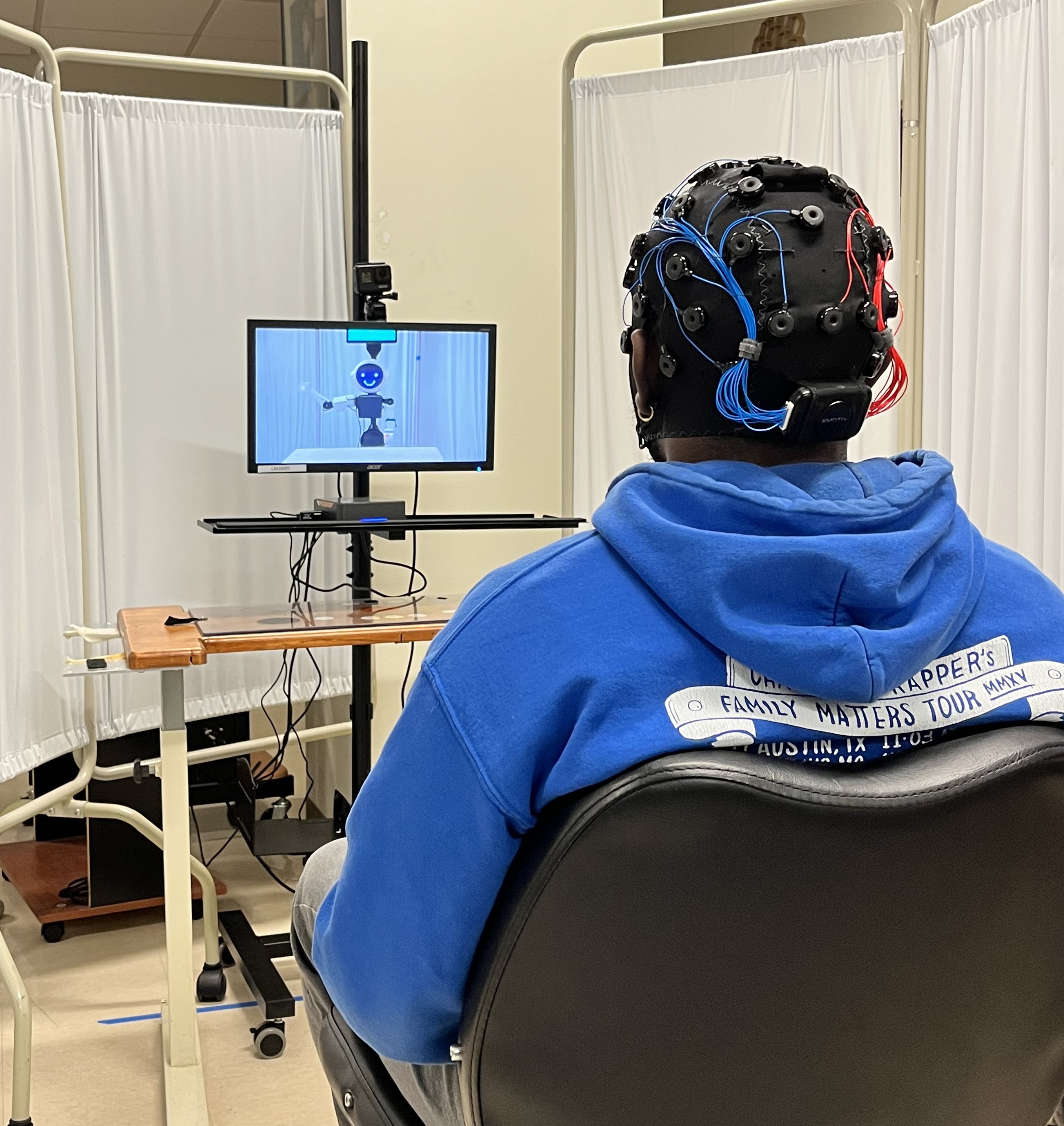}%
  }}
  \cr\noalign{\hfill}
  \hbox{\subcaptionbox*{(C)}[.35\linewidth]{%
    \includegraphics[width=\linewidth]{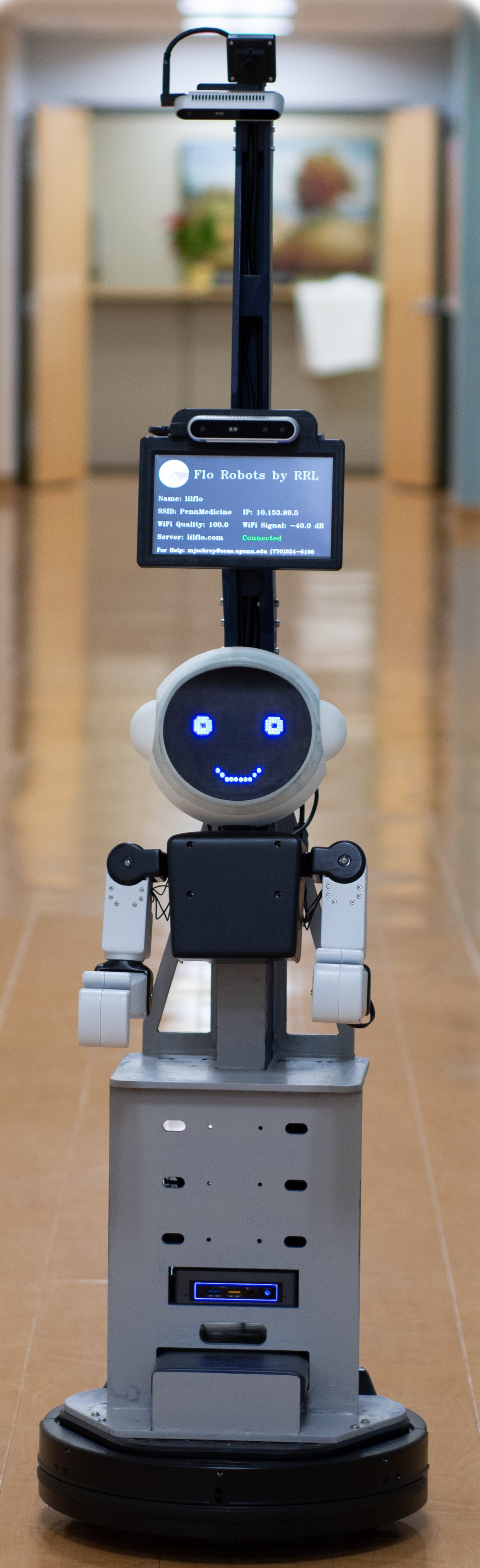}%
  }}\cr
}
    \caption{(A) The human actor alongside the humanoid social robot Flo, in their respective configurations used for AO task demonstrations (B) One of the subjects participating in the experiment, observing a video clip of "robot" condition (C) The humanoid social robot Flo, mounted on its custom telepresence base. }\label{images}
\end{figure}

\begin{figure*}[h]
    \centering
    \includegraphics[width=0.8\textwidth]{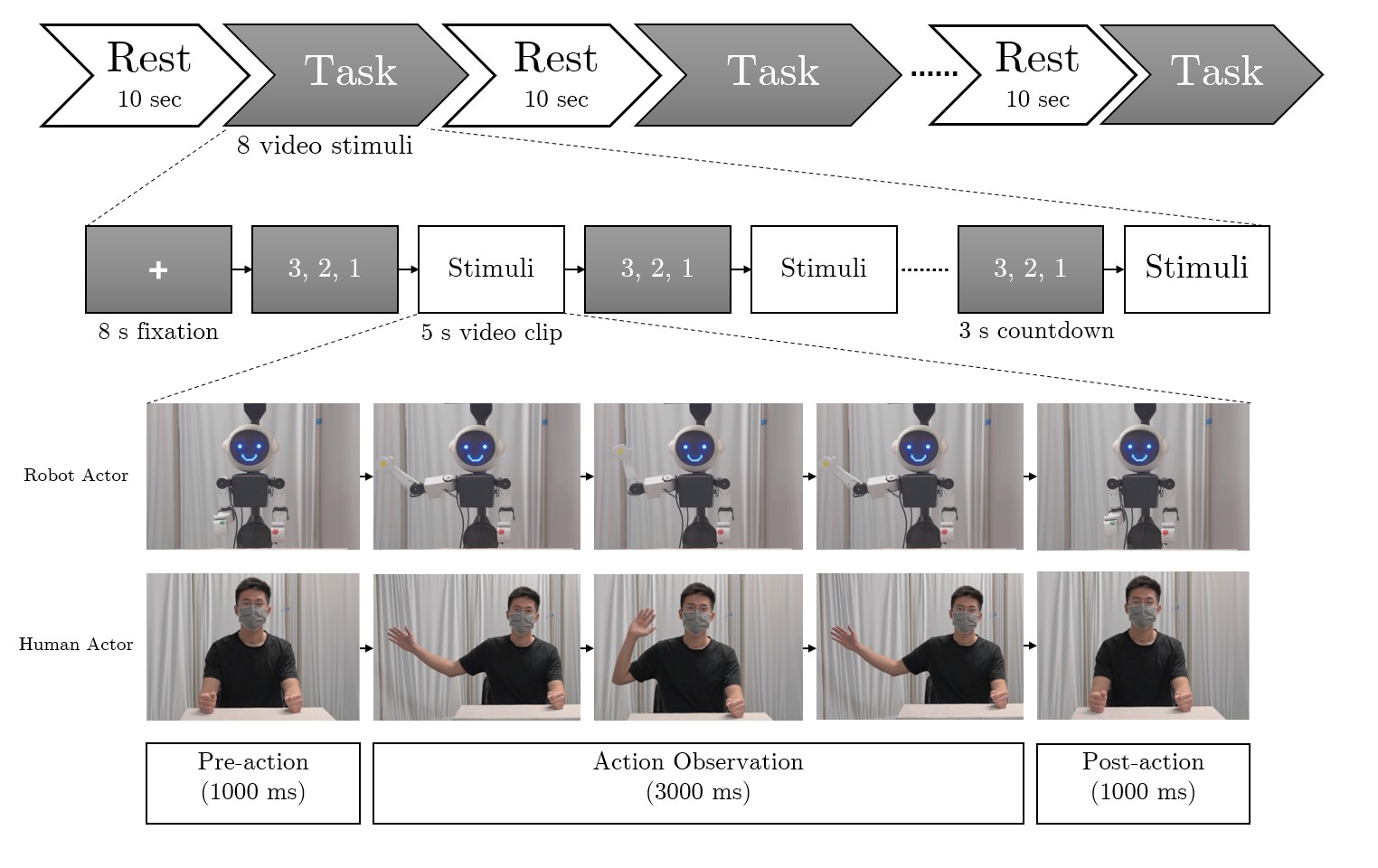}
    \caption{Block design of the video stimuli administration. There were a total of 12 blocks of AO task, each block consisted of 8000 ms of fixation cross and a sequence of eight 5000-ms video stimuli of the same AO condition. }
    \label{AO_Task}
\end{figure*}

The stimuli for both experiments comprised eight 5000-ms videos, depicting various uni-manual arm movements such as air punching, backward-forward arm swing, lateral arm swing, reaching forward, reaching to the side, covering eye, touching head, and waving. 
Each video clip showed in sequence: 1000 ms pre-action interval, in which the actor's arm is either in resting position or moving to starting position of the main action; 3000 ms action observation interval; 1000 ms post-action in which the actor's arm returns to the resting position. 
These actions were performed by two types of actors, human and social robot, utilizing both their left and right hands, and were recorded in an allocentric perspective (Fig. \ref{images}A).
A total of thirty-two videos were captured and subsequently categorized into four distinct experimental conditions. These conditions were determined by the combination of two different actors (human, robot) and two observed hand (right, left): (1) human-left; (2) human-right; (3) robot-left; (4) robot-left.
The administration of stimuli followed a block design design and was counterbalanced to ensure robust experimental control (Fig. \ref{AO_Task}). Each stimulus video was repeated three times and organized into blocks consisting of eight stimuli of the same condition (human-left, human-right, robot-left, robot-left). The presentation of stimuli consisted of a total of 12 blocks. To alleviate possible interference during baseline EEG acquisition due to the momentarily decrease in the subject's attention or the influx of miscellaneous waves, an 8000-ms fixation cross period was used at the beginning of each block, in which participants were instructed to fixate on a white cross on a black screen \cite{kim2021eeg}. The fixation cross period was followed by the sequential presentation of eight randomized video stimuli of the same condition, which was interleaved with countdown intervals of 3000 ms. 
Visual stimuli were administered using open-source software PsychoPy  \cite{peirce2019psychopy2}.

The participants were positioned 150 cm away from a 24-inch LCD computer screen (Fig. \ref{images}B). They were asked to keep their hands still on their lap while observing the video stimuli from the four experimental conditions mentioned above.
Before starting the EEG recording, participants underwent a brief training session in which they were presented with the fixation-cross screen and two videos, each from the "robot" and "human" conditions. They were instructed to fixate on the center cross and observe the video clips.

\subsection{Robot Demonstrator}
The robotic system employed for the illustration of AO tasks encompasses a bespoke social robot named Flo (formerly known as Lil'Flo) designed and fabricated in the Rehab Robotics Lab at the University of Pennsylvania \cite{sobrepera2021DesignLilFlo} (Fig. \ref{images}C). Flo is a humanoid social robot, possessing both an expressive facial interface and articulate upper limbs \cite{sobrepera2019DesigningEvaluatingFace, sobrepera2019DesigningArmsLil}. It has been specifically engineered for the purpose of illustrating upper extremity rehabilitation tasks in preceding experimental investigations. It is part of a telerehabilitation system that allows a remote operator/clinician to interact with patients and use the robot as an aid to demonstrate rehabilitation activities. In our previous studies, it compared favorably to classical telepresence methods and, thus, was considered suitable to compare against a human to study MNS activation during AO \cite{sobrepera2021PerceivedUsefulnessSocial, Thesis}.

\subsection{EEG Recording and Data Analysis}
The EEG recording in this study used the Emotiv EPOC Flex EEG system (Emotiv Inc.), which comprises 32 Ag-AgCl electrodes. Each electrode was equipped with saline soaked felt pads and affixed to an EasyCap\texttrademark (Herrsching, Germany), which was configured based on the international 10–20 system for electrode placement. There are two common-mode sensors located at left and right mastoids for referencing during EEG recordings. The Flex system featured built-in EEG data pre-processing functionalities, incorporating a high-pass filter set at 0.2 Hz and a low-pass filter at 45 Hz. Additionally, the system performed digitization at a rate of 1,024 Hz and applied a 5th order digital sine filter for filtering, followed by downsampling to 128 Hz. The acquisition of EEG data was performed using the Emotiv Pro software.
The exported EEG data underwent further off-line pre-processing and analysis using MATLAB (The MathWorks, Inc., Natick, MA, USA) and the EEGLAB toolbox \cite{delorme2004eeglab}. 
Raw data was cleaned using both an automated algorithm (clean\_rawdata plugin) and visual inspection to reject bad channels (i.e., channels with sudden shifts, excessive high-frequency noise, and excessive voltage amplitude compared to other channels) and then re-referenced to the average following the interpolation of the removed channels. Artifacts such as ocular, cardiac or muscular were labeled and removed using Independent Component Analysis (ICA). The fixation cross period (duration: 8000 ms) and the stimuli epochs (duration: 5000 ms) were extracted from two channels of interest (C3 and C4) to examine sensorimotor mu oscillations. 
To alleviate the variability of EEG power due to possible EEG interference during the long interval of fixation cross, a baseline selection step was performed.
For each fixation cross period, 1000-ms baseline were identified by computing the correlation of averaged power spectrum across all segments between the 8000-ms fixation cross and each of its eight 1000-ms time windows, separately for three frequency bands: Theta (4-8 Hz), mu (8-13 Hz), and beta (13-30 Hz). The 1000 ms time window that had the highest value of averaged correlation coefficient across the frequency bands served as the baseline for the subsequent single-subject analysis for each channel. 

For each participant and condition, the time-frequency analysis for the stimuli epochs was performed between 4 and 30 Hz, with a resolution of 0.5 Hz \cite{delorme2004eeglab, angelini2018perspective}.
To take into account individual differences in overall EEG power, the spectral data were normalized by dividing each value at each time-frequency point by the average spectral power of the baseline at the corresponding frequency. The resulting data were expressed as absolute power ratios relative to the baseline activity.
For each participant, condition, channel, and frequency band of interest, the ratio data were averaged over time. Since the absolute power ratio data had a non-normal distribution, a log10 transformation was performed on each absolute power ratio value. A negative log10 ratio indicated the event-related desynchronization (ERD), while a positive log10 ratio indicated a relative EEG power increase or event-related synchronization (ERS). 
Correlation coefficients were computed to assess the relationship between EEG power ratios for each pair of experimental conditions for each participant, using Pearson's correlation coefficient. Least squares plots were generated to visually represent the correlation.
Event-related spectral perturbations during AO were also analyzed using visualization of power spectral density and baseline-corrected power ratio for each participant and channel.

%=====================================================
\section{Results}
\begin{figure}[h]
\centering
        \resizebox{\linewidth}{!}{\input{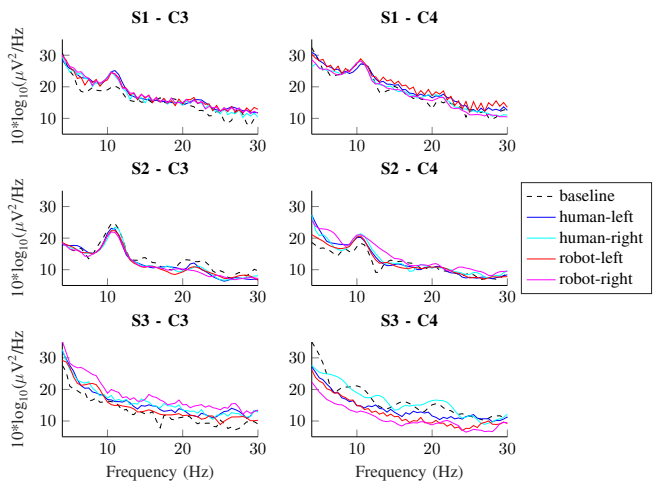}}
        \caption{Power spectra of baseline and stimuli epochs for each participant and channel. }\label{fig:psd}
\end{figure}

A peak within the mu band (8-13 Hz) was visible during the baseline and stimulus epochs at both the C3 and C4 channels of subjects 1 and 2, but was only visible in the C3 channel of subject 3 (Fig. \ref{fig:psd}). 
Two peaks centered around 15 and 21 Hz within the beta band were visible during baseline in the power spectra at the C3 channel of subject 2 and at the C4 channel of subjects 1 and 3. For the power spectra during AO, the peak centered around 21 Hz was visible but not significant at both channels of subjects 1 and 2, and a peak in the power spectrum of subject 3 at C4 was observable for the "human-right" condition.

\begin{figure}[h]
\vspace{5pt}
\centering
        \resizebox{\linewidth}{!}{% This file was created by matlab2tikz.
%
%The latest updates can be retrieved from
%  http://www.mathworks.com/matlabcentral/fileexchange/22022-matlab2tikz-matlab2tikz
%where you can also make suggestions and rate matlab2tikz.
%
\definecolor{mycolor1}{rgb}{1.00000,0.00000,1.00000}%
\begin{tikzpicture}

\begin{axis}[%
width=1.683in,
height=0.739in,
at={(0.758in,3.107in)},
scale only axis,
point meta min=-0.8,
point meta max=0.8,
axis on top,
xmin=63.4782608695652,
xmax=3436.52173913043,
xlabel style={font=\color{white!15!black}},
xlabel={},
ymin=4,
ymax=30,
ylabel style={font=\color{white!15!black}},
ylabel={Frequency (Hz)},
axis background/.style={fill=white},
title style={font=\bfseries},
title={S1 - C3},
colormap/jet,
colorbar,
colorbar style={title={Log10 Power Ratio}}
]
\addplot [forget plot] graphics [xmin=59.2513893429226, xmax=3440.74861065708, ymin=3.75, ymax=30.25] {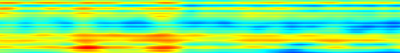};
\addplot [color=mycolor1, dashed, forget plot]
  table[row sep=crcr]{%
63.4782608695652	8\\
3436.52173913043	8\\
};
\addplot [color=mycolor1, dashed, forget plot]
  table[row sep=crcr]{%
63.4782608695652	13\\
3436.52173913043	13\\
};
\addplot [color=mycolor1, dashed, forget plot]
  table[row sep=crcr]{%
63.4782608695652	20\\
3436.52173913043	20\\
};
\end{axis}

\begin{axis}[%
width=1.683in,
height=0.739in,
at={(3.448in,3.107in)},
scale only axis,
point meta min=-0.8,
point meta max=0.8,
axis on top,
xmin=63.4782608695652,
xmax=3436.52173913043,
xlabel style={font=\color{white!15!black}},
xlabel={ },
ymin=4,
ymax=30,
ylabel style={font=\color{white!15!black}},
ylabel={  },
axis background/.style={fill=white},
title style={font=\bfseries},
title={S1 - C4},
colormap/jet,
colorbar,
colorbar style={title={Log10 Power Ratio}}
]
\addplot [forget plot] graphics [xmin=59.2513893429226, xmax=3440.74861065708, ymin=3.75, ymax=30.25] {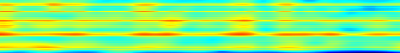};
\addplot [color=mycolor1, dashed, forget plot]
  table[row sep=crcr]{%
63.4782608695652	8\\
3436.52173913043	8\\
};
\addplot [color=mycolor1, dashed, forget plot]
  table[row sep=crcr]{%
63.4782608695652	13\\
3436.52173913043	13\\
};
\addplot [color=mycolor1, dashed, forget plot]
  table[row sep=crcr]{%
63.4782608695652	20\\
3436.52173913043	20\\
};
\end{axis}

\begin{axis}[%
width=1.683in,
height=0.739in,
at={(0.758in,1.794in)},
scale only axis,
point meta min=-0.8,
point meta max=0.8,
axis on top,
xmin=63.4782608695652,
xmax=3436.52173913043,
xlabel style={font=\color{white!15!black}},
xlabel={ },
ymin=4,
ymax=30,
ylabel style={font=\color{white!15!black}},
ylabel={Frequency (Hz)},
axis background/.style={fill=white},
title style={font=\bfseries},
title={S2 - C3},
colormap/jet,
colorbar,
colorbar style={title={Log10 Power Ratio}}
]
\addplot [forget plot] graphics [xmin=59.2513893429226, xmax=3440.74861065708, ymin=3.75, ymax=30.25] {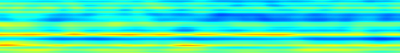};
\addplot [color=mycolor1, dashed, forget plot]
  table[row sep=crcr]{%
63.4782608695652	8\\
3436.52173913043	8\\
};
\addplot [color=mycolor1, dashed, forget plot]
  table[row sep=crcr]{%
63.4782608695652	13\\
3436.52173913043	13\\
};
\addplot [color=mycolor1, dashed, forget plot]
  table[row sep=crcr]{%
63.4782608695652	20\\
3436.52173913043	20\\
};
\end{axis}

\begin{axis}[%
width=1.683in,
height=0.739in,
at={(3.448in,1.794in)},
scale only axis,
point meta min=-0.8,
point meta max=0.8,
axis on top,
xmin=63.4782608695652,
xmax=3436.52173913043,
xlabel style={font=\color{white!15!black}},
xlabel={ },
ymin=4,
ymax=30,
ylabel style={font=\color{white!15!black}},
ylabel={  },
axis background/.style={fill=white},
title style={font=\bfseries},
title={S2 - C4},
colormap/jet,
colorbar,
colorbar style={title={Log10 Power Ratio}}
]
\addplot [forget plot] graphics [xmin=59.2513893429226, xmax=3440.74861065708, ymin=3.75, ymax=30.25] {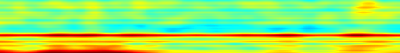};
\addplot [color=mycolor1, dashed, forget plot]
  table[row sep=crcr]{%
63.4782608695652	8\\
3436.52173913043	8\\
};
\addplot [color=mycolor1, dashed, forget plot]
  table[row sep=crcr]{%
63.4782608695652	13\\
3436.52173913043	13\\
};
\addplot [color=mycolor1, dashed, forget plot]
  table[row sep=crcr]{%
63.4782608695652	20\\
3436.52173913043	20\\
};
\end{axis}

\begin{axis}[%
width=1.683in,
height=0.739in,
at={(0.758in,0.481in)},
scale only axis,
point meta min=-0.8,
point meta max=0.8,
axis on top,
xmin=63.4782608695652,
xmax=3436.52173913043,
xlabel style={font=\color{white!15!black}},
xlabel={Time (ms)},
ymin=4,
ymax=30,
ylabel style={font=\color{white!15!black}},
ylabel={Frequency (Hz)},
axis background/.style={fill=white},
title style={font=\bfseries},
title={S3 - C3},
colormap/jet,
colorbar,
colorbar style={title={Log10 Power Ratio}}
]
\addplot [forget plot] graphics [xmin=59.2513893429226, xmax=3440.74861065708, ymin=3.75, ymax=30.25] {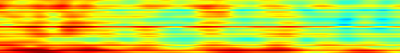};
\addplot [color=mycolor1, dashed, forget plot]
  table[row sep=crcr]{%
63.4782608695652	8\\
3436.52173913043	8\\
};
\addplot [color=mycolor1, dashed, forget plot]
  table[row sep=crcr]{%
63.4782608695652	13\\
3436.52173913043	13\\
};
\addplot [color=mycolor1, dashed, forget plot]
  table[row sep=crcr]{%
63.4782608695652	20\\
3436.52173913043	20\\
};
\end{axis}

\begin{axis}[%
width=1.683in,
height=0.739in,
at={(3.448in,0.481in)},
scale only axis,
point meta min=-0.8,
point meta max=0.8,
axis on top,
xmin=63.4782608695652,
xmax=3436.52173913043,
xlabel style={font=\color{white!15!black}},
xlabel={Time (ms)},
ymin=4,
ymax=30,
ylabel style={font=\color{white!15!black}},
ylabel={  },
axis background/.style={fill=white},
title style={font=\bfseries},
title={S3 - C4},
colormap/jet,
colorbar,
colorbar style={title={Log10 Power Ratio}}
]
\addplot [forget plot] graphics [xmin=59.2513893429226, xmax=3440.74861065708, ymin=3.75, ymax=30.25] {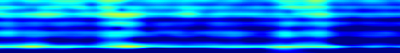};
\addplot [color=mycolor1, dashed, forget plot]
  table[row sep=crcr]{%
63.4782608695652	8\\
3436.52173913043	8\\
};
\addplot [color=mycolor1, dashed, forget plot]
  table[row sep=crcr]{%
63.4782608695652	13\\
3436.52173913043	13\\
};
\addplot [color=mycolor1, dashed, forget plot]
  table[row sep=crcr]{%
63.4782608695652	20\\
3436.52173913043	20\\
};
\addplot [color=mycolor1, dashed, forget plot]
  table[row sep=crcr]{%
63.4782608695652	8\\
3436.52173913043	8\\
};
\addplot [color=mycolor1, dashed, forget plot]
  table[row sep=crcr]{%
63.4782608695652	13\\
3436.52173913043	13\\
};
\addplot [color=mycolor1, dashed, forget plot]
  table[row sep=crcr]{%
63.4782608695652	20\\
3436.52173913043	20\\
};
\addplot [color=mycolor1, dashed, forget plot]
  table[row sep=crcr]{%
63.4782608695652	8\\
3436.52173913043	8\\
};
\addplot [color=mycolor1, dashed, forget plot]
  table[row sep=crcr]{%
63.4782608695652	13\\
3436.52173913043	13\\
};
\addplot [color=mycolor1, dashed, forget plot]
  table[row sep=crcr]{%
63.4782608695652	20\\
3436.52173913043	20\\
};
\addplot [color=mycolor1, dashed, forget plot]
  table[row sep=crcr]{%
63.4782608695652	8\\
3436.52173913043	8\\
};
\addplot [color=mycolor1, dashed, forget plot]
  table[row sep=crcr]{%
63.4782608695652	13\\
3436.52173913043	13\\
};
\addplot [color=mycolor1, dashed, forget plot]
  table[row sep=crcr]{%
63.4782608695652	20\\
3436.52173913043	20\\
};
\addplot [color=mycolor1, dashed, forget plot]
  table[row sep=crcr]{%
63.4782608695652	8\\
3436.52173913043	8\\
};
\addplot [color=mycolor1, dashed, forget plot]
  table[row sep=crcr]{%
63.4782608695652	13\\
3436.52173913043	13\\
};
\addplot [color=mycolor1, dashed, forget plot]
  table[row sep=crcr]{%
63.4782608695652	20\\
3436.52173913043	20\\
};
\addplot [color=mycolor1, dashed, forget plot]
  table[row sep=crcr]{%
63.4782608695652	8\\
3436.52173913043	8\\
};
\addplot [color=mycolor1, dashed, forget plot]
  table[row sep=crcr]{%
63.4782608695652	13\\
3436.52173913043	13\\
};
\addplot [color=mycolor1, dashed, forget plot]
  table[row sep=crcr]{%
63.4782608695652	20\\
3436.52173913043	20\\
};
\end{axis}
\end{tikzpicture}%
}
        \caption{Event-related spectral perturbations (ERSP) of C3 and C4 channel of three participants averaged across all conditions, showing the time-frequency presentation of log10 power ratio relative to baseline activity for the stimuli epoch. The onset of the video stimuli is at 0 ms. Pink dashed lines indicate the analyzed frequency ranges, including theta (4-8 Hz), mu (8–13 Hz) and beta (13–30 Hz).}\label{fig:ersp}.
\end{figure}
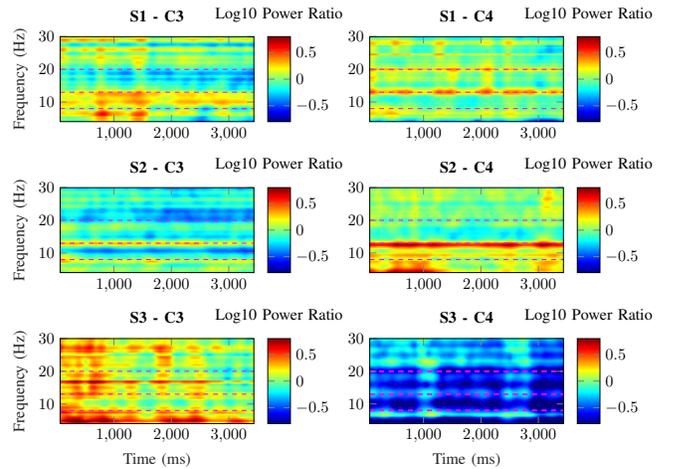

\begin{figure}[h]
% \vspace{5pt}
    \centering
    \begin{subfigure}[b]{0.5\textwidth}
        \resizebox{\linewidth}{!}{\input{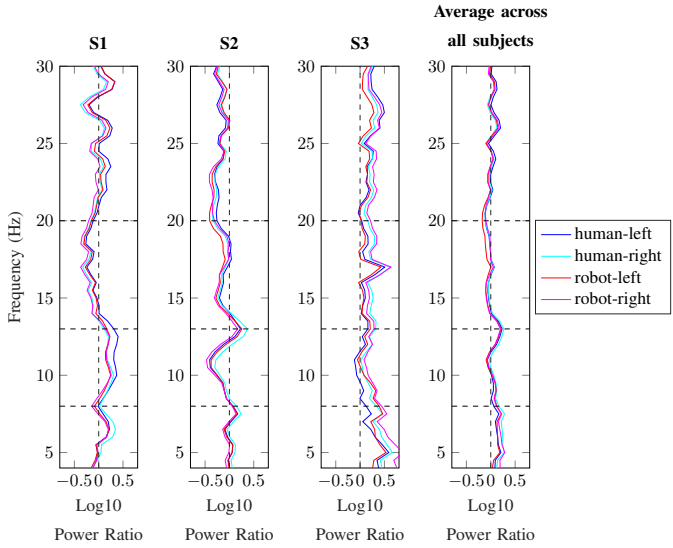}}
        \caption{}
        \label{fig:C3_allersp}
    \end{subfigure}

    \begin{subfigure}[b]{0.5\textwidth}
    \centering
        \resizebox{\linewidth}{!}{\input{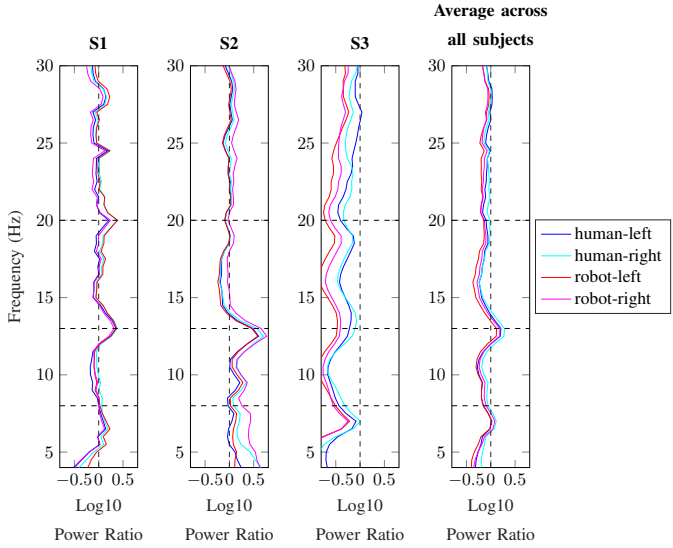}}
        \caption{}
        \label{fig:C4_allersp}
    \end{subfigure}
        \caption{Averaged log10 power ratio of (a) C3 electrode and (b) C4 electrode for each frequency during AO period across all conditions for each participant and the grand-averaged log10 power ratio across all participants. Black horizontal dashed lines indicate the analyzed frequency ranges, including theta (4-8 Hz), mu (8–13 Hz) and beta (13–30 Hz).}
\end{figure}

Figure \ref{fig:ersp} shows the time-frequency decomposition of ERSP over the stimulus epoch averaged across all conditions of AO. The ERSP showed the suppression of mu rhythm at the C3 channel of subject 2 and the C4 channel of subject 3 (Fig. \ref{fig:ersp}). Event-related synchronizations (ERS) of the mu band at the C3 channel were observed for both subjects 1 and 3. For all participants and channels, the ERD and ERS patterns of the mu and beta bands were fairly consistent during the stimuli epoch, except for the early ERS observed from the onset of the stimuli to approximately 1000 ms in the C3 channel of subject 3. While only C4 channel of subject 3 observed event-related desynchronization (ERD) in the beta band, beta desynchronizations were visible at the C3 channel of both subjects 1 and 2.
Theta synchronizations were observed at the C3 channel of subjects 1 and 3 and the C4 channel of subject 2.
These phenomena were also seen in Figures \ref{fig:C3_allersp} and \ref{fig:C4_allersp}, which display the averaged log10 power ratio across the stimulus period for each frequency, AO condition and participant at C3 and C4 channel respectively.
The ERD and ERP patterns described above were true for all AO conditions for the three participants.
Interestingly, the right sensorimotor region of subject 3 showed substantial ERSs across all frequency bands, specifically theta band (4-7 Hz), mu band (8-13 Hz), and beta band (13-23 Hz), and the ERD effect of "robot" conditions of AO were stronger than the "human" conditions (Fig. \ref{fig:C4_allersp}). 

\begin{figure}[!]
\vspace{5pt}
    \centering
    \begin{subfigure}[b]{0.5\textwidth}
        \resizebox{\linewidth}{!}{\input{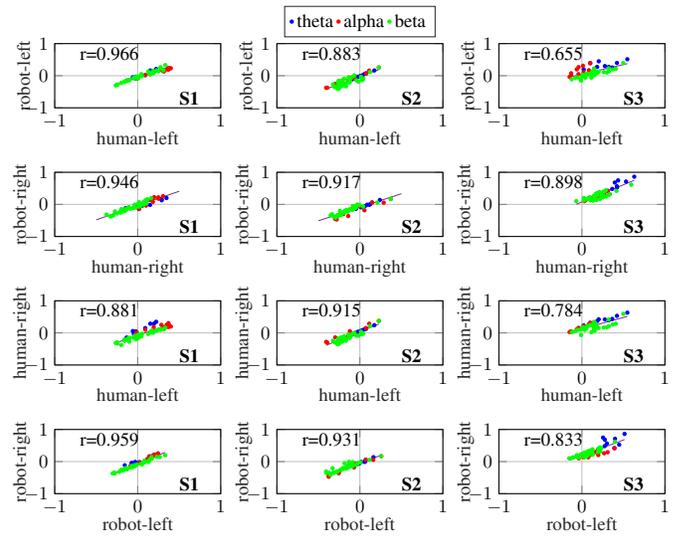}}
        \caption{}
        \label{fig:C3_corr}
    \end{subfigure}

    \begin{subfigure}[b]{0.5\textwidth}
    \centering
        \resizebox{\linewidth}{!}{\input{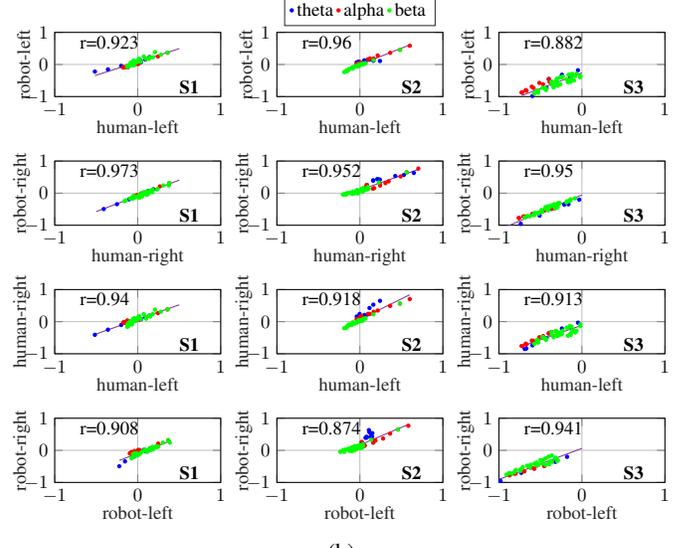}}
        \caption{}
        \label{fig:C4_corr}
    \end{subfigure}
        \caption{Correlation scatterplot of the log10 power ratio at each frequency in 4-30 Hz range between pairs of conditions for each participant at (a) C3 electrode and (b) C4 electrode. Least squared lines illustrate the linear trend between the EEG power ratios of the paired conditions, with the correlation coefficient displayed on the plot. Points denoting values of frequency within each of three frequency bands have different colors.}
\end{figure}

Figures \ref{fig:C3_corr} and \ref{fig:C4_corr} show correlations between the power ratio results of observing human and robot actors using left and right arms at channels C3 and C4. Overall, the correlations between pairs of conditions were positive and ranged from moderate to strong.  For the C3 channel, the correlation coefficients of the log10 power ratio between pairs of conditions showed strong and positive linear relationships for subjects 1 and 2 ($r>0.88$), and moderate for the corresponding pairs of conditions of subject 3 (Fig. \ref{fig:C3_corr}) with the weaker correlation between the "robot-left" and "human-left" conditions ($r=0.655$). 
There were more frequencies within the mu band (red points) in the left sensorimotor region that show power synchronisation from observing human and left-hand actions than from observing robot and left-hand actions (Fig. \ref{fig:C3_corr}). Similar to the C3 channel, the log10 power ratios between pairs of conditions in C4 were strong and positive for all three participants ($r>0.87$) (Fig. \ref{fig:C4_corr}).

% \begin{figure}[h]
% \vspace{5pt}
% \centering
%         \resizebox{\linewidth}{!}{\input{C3_allersp}}
%         \caption{Averaged log10 power ratio of C3 electrode for each frequency during AO period across all conditions for each participant and the grand-averaged log10 power ratio across all participants. Black horizontal dashed lines indicate the analyzed frequency ranges, including theta (4-8 Hz), mu (8–13 Hz) and beta (13–30 Hz).}\label{fig:C3_allersp}
% \end{figure}

% \begin{figure}[!]
% \centering
%         \resizebox{\linewidth}{!}{\input{C4_allersp}}
%         \caption{Averaged log10 power ratio of C4 electrode for each frequency during AO period across all conditions for each participant and the grand-averaged log10 power ratio across all participants. Black horizontal dashed lines indicate the analyzed frequency ranges, including theta (4-8 Hz), mu (8–13 Hz) and beta (13–30 Hz).}\label{fig:C4_allersp}
% \end{figure}

% \begin{figure}[h]
% \vspace{5pt}
%     \centering
%         \resizebox{\linewidth}{!}{\input{C3_corr}}
%         \caption{Correlation scatterplots of the log10 power ratio at each frequency in 4-30 Hz range between pairs of conditions for each participant at C3 electrode. Least squared lines illustrate the linear trend between the EEG power ratios of the paired conditions, with the correlation coefficient displayed on the plot. Points denoting values of frequency within each of three frequency bands have different colors.}\label{fig:C3_corr}
% \end{figure}

% \begin{figure}[!]
%     \centering
%         \resizebox{\linewidth}{!}{\input{C4_corr}}
%         \caption{Correlation scatterplot of the log10 power ratio at each frequency in 4-30 Hz range between pairs of conditions for each participant at C4 electrode. Least squared lines illustrate the linear trend between the EEG power ratios of the paired conditions, with the correlation coefficient displayed on the plot. Points denoting values of frequency within each of three frequency bands have different colors.}\label{fig:C4_corr}
% \end{figure}

\section{Discussion}
The present EEG case study with three right-handed healthy participants aimed to investigate neural correlates during the observation of actions performed by human and robot actor, using left and right hand. We sought to employ an EEG-based approach to explore the neurophysiology of human-robot interaction, which provides a foundation for long-term future research comparing the neural responses to human and robot actors in AO and, furthermore, exploring the neurorehabilitation effects of using social robot augmented telepresence (SRAT) for stroke patients. The EEG responses during AO were analyzed using ERSP at the left and right sensorimotor regions. 
Clearly it is premature to generalize these preliminary results, they provide insights into the feasibility of the employed EEG methodology and measures such as power suppression of mu and beta rhythms to investigate neural responses during observation of both human and robot actions. The correlation results imply that neural responses seen in observing robot actions may not be different from those seen in observing human actions. The study also showed unique neural signatures among three participants, highlighting the importance of considering individual differences in AO studies, particularly in the context of human-robot interaction. 

The sensorimotor cortex is active not only during movement execution, but also when observing movements and actions of others, through activation of the Mirror Neuron System (MNS) \cite{rizzolatti2001neurophysiological}. Desynchronization in mu band is thought to be an indicator of MNS activation \cite{hobson2017interpretation}. The mu rhythm is thought to be spontaneously generated by sensory motor nerves during rest, indicated by a peak in the mu band (8-13 Hz) in the EEG power spectra. 
This phenomenon was observed in sensorimotor regions in the power spectra of two of the three participants when observing robot and human actors. 
The varied patterns seen in mu rhythm of subject 3 might suggest the individual variability of mu band, and the absence of mu rhythm could indicate the states of heightened attention even during fixation cross which served as baseline.
% This could also highlight the use of Independent Component Analysis (ICA) to identify mu rhythms (i.e., mu components) which can address the limitations of the conventional channel-based measures of spectral power used in this study \cite{jenson2020application}. 
There is still uncertainty regarding whether subject 3 genuinely lack a mu rhythm or if we were unable to detect mu activity in these cases due to factors such as low signal-to-noise ratio \cite{haegens2014inter}. More subjects are needed to determine the consistency of this phenomenon. 

Mu ERD in the sensorimotor cortex is considered a neural signature of the engagement of motor-related processes during action perception, indicating inhibition of motor neurons during AO \cite{rizzolatti2001neurophysiological}.
In our case study, only one participant (subject 2) showed mu ERD at C3 channel demonstrating left-lateralized primary motor response during AO for both robot and human AO, and the other two participants were right-lateralized to different degrees. These results again suggest the individual differences in motor activity during AO of both human and robots. These patterns were consistent for both "right" and "left" hand actions. This is in line with a transcranial magnetic stimulation (TMS) study by Sartori et al. finding that right-handed individuals exhibited left-lateralized primary motor responses to observations of left and right grasping actions from an allocentric perspective \cite{sartori2013motor}. These findings are inconsistent with a common belief that representation of observations is in a spatially compatible (i.e., mirror-like) manner, specifically, the neural responses to observed right-hand actions are mainly localized in the right hemisphere, while the neural responses to observed left-hand actions are mainly localized in the left hemisphere. \cite{hesse2009end, vingerhoets2012influence}.

The neural responses to "human" and "robot" conditions of all participants in both channels had similar patterns of ERDs and ERSs throughout the frequency range of 4-30 Hz. In addition, nearly all correlations in the log10 power ratio at the three frequency bands between pairs of "human" and "robot" conditions for each participant and channel show strong and positive relationships. This result appears consistent with previous EEG studies showing that the observations of human and robot actions, both with and without object manipulation, induced equivalent mu rhythm suppression \cite{oberman2007eeg, urgen2013eeg}. 
The exception is the results at the C4 channel of subject 3 which displayed substantial difference in ERD between "human" (red and magenta curves) and "robot" (blue and cyan curves) conditions across beta rhythm (13-30 Hz), specifically more beta power suppression during "robot" AO conditions. Similar findings were also observed in an fMRI study, in which both parietal and premotor regions responded significantly more during observation of grasping actions performed by a robot compared to a human \cite{kupferberg2018fronto}.
Responses of action observation network are sensitive to the goals of observed actions, such as target objects and overarching purposes \cite{errante2019parieto, kemmerer2021modulates}. The eight actions completed by our robot and human actor did not involve manipulation of objects. This highlights the need for our future study to explore neural responses to robot actions that involve object manipulation.

\subsection*{Limitations}
Due to the small sample size, we are not able to draw any conclusive and generalizable findings. However, the exploratory nature of the investigation allows us to examine the subject-specific variability and patterns of power spectra and event-related spectral perturbations. 
% We were able to verify the feasibility of EEG-based approach to explore the neural responses to different conditions of action observation and their correlation. 
Some patterns of power spectra and ERSP are consistent with findings in literature, motivating us to employ this approach for further research with larger scale. 
% Another limitation lies in the experiment design, in that the fixation cross segments expected to serve as the baseline were too long (8000 ms), requiring additional preprocessing (i.e., baseline selection). Future research will adopt shorter baseline periods (e.g., 500-1000 ms) preceding each stimulus to directly calculate ERSP, facilitated by EEGLAB, and illustrate temporal variations in EEG power from baseline to stimulus onset. 
Additionally, as suggested above, the traditional channel-based method for identifying mu-alpha power might be influenced by residual artifacts and volume conduction, as well as our inability to separate mixed signals due to low density EEG and dependence on re-referencing techniques. 
Future research should implement ICA to analyze mu and beta components, which can provide a more robust approach for mapping neural activity to behavior,  surpassing the limitations associated with the traditional channel-based method \cite{jenson2020application}.

\section{Conclusion}
The study provides preliminary evidence of the feasibility and potential of using EEG to explore the single-subject patterns and variability in neural dynamics of AO that involves human and robot actors. Expected patterns of event-related spectral perturbation were observed in the mu-alpha and beta bands (power suppression) across all participants. One participant showed prominent difference in ERD of the beta band between "human" and "robot" conditions of AO, specifically stronger ERD effects of the "robot" conditions. 
Nearly all correlations in the log10 power ratio at the three frequency bands between pairs of conditions for each participant and channel are strong and positive, suggesting consistent ERSP patterns across conditions and common processes involved in perceiving and processing observed actions.
Future studies should further analyze EEG data, including ERSP and functional coherence, across all channels, and involve stroke patients to explore neurocognitive mechanisms in robot-assisted AO therapy for stroke rehabilitation.

\bibliography{refs.bib}
\end{document}